\documentclass[a4paper]{article}
\usepackage{amsmath,amssymb}
\usepackage[export]{adjustbox}
\usepackage{graphicx}

\begin{document}

\title{On galaxy rotation curves from a continuum mechanics approach to modified gravity}

\author{
Christian G.~B\"ohmer\footnote{c.boehmer@ucl.ac.uk}, 
Nicola Tamanini\footnote{nicola.tamanini@cea.fr}, and 
Matthew Wright\footnote{matthew.wright.13@ucl.ac.uk}\\[1ex]
${}^{*\ddagger}$Department of Mathematics, University College London\\
Gower Street, London, WC1E 6BT, UK\\[1ex]
${}^{\dagger}$Institut de Physique Th{\'e}orique, CEA-Saclay\\ F-91191 Gif-sur-Yvette, France}

\date{New version of \today}

\maketitle

\begin{abstract}
We consider a modification of General Relativity motivated by the treatment of anisotropies in Continuum Mechanics. The Newtonian limit of the theory is formulated and applied to galactic rotation curves. By assuming that the additional structure of spacetime behaves like a Newtonian gravitational potential for small deviations from isotropy, we are able to recover the Navarro-Frenk-White profile of dark matter halos by a suitable identification of constants. We consider the Burkert profile in the context of our model and also discuss rotation curves more generally.
\end{abstract}

\section{Introduction}

Astrophysics and cosmology are faced with two severe theoretical difficulties, that can be summarised as the dark energy and the dark matter problems. To address the first, one could, in principle, accept the cosmological constant to be a small fundamental constant of physics and neglect issues arising from particle physics. In fact the cosmological constant is consistent with all observations to date, see~\cite{Ade:2015xua}. However it appears that there is no obvious route to escape the need for dark matter. We cannot `solve' the dark matter problem by adding an additional constant to physics. The main experimental evidence for the existence of dark matter comes from the behaviour of the galactic rotation curves, first observed by Rubin~\cite{Rubin:1980zd}, and the mass discrepancy in galactic clusters. Both suggest there exits some form of matter at galactic and extra-galactic scales which only interacts very weakly with normal matter, its main interaction being via the gravitational force~\cite{Clowe:2006eq}. On cosmological scales, recent Planck data~\cite{Ade:2015xua} puts very tight constraints on the amounts of dark matter and dark energy, respectively, confirming the presence of these components in the universe. 

Galactic rotation curves of spiral galaxies give strong evidence for the presence of some additional form of matter. One can observe neutral hydrogen clouds at large distances from the centre of the galaxy where the Newtonian gravitational field is weak. These clouds lie way outside the luminous part of the galaxy. Observations show that these clouds are moving at approximately constant tangential velocity $v_{\rm tg}$. Newton's law of gravity together with the centrifugal force yield the well known relation
\begin{align*}
  \frac{v_{\rm tg}^2}{r^2} = \frac{GM}{r^3}.
\end{align*}
In order for parts of the outer galaxy to move with approximately constant tangential velocity would require the mass of this region to grow with $r$. This is in stark contrast to observations which show that these regions contain little luminous matter. Thus, an additional (dark, since we cannot see it) matter component is required to explain this behaviour. There exists a plethora of dark matter models, see e.g.~\cite{Arrenberg:2013rzp}, having their roots either in particle physics or modified gravity.

In~\cite{Boehmer:2013ss} a new approach to modifying general relativity was formulated. It is based on ideas well known in Continuum Mechanics. For similar works related to dark energy see also~\cite{Pearson:2014iaa}, while in the context of cosmological dark matter, continuum mechanics inspired models were studied in~\cite{Bucher:1998mh,Celoria:2017bbh}. These approaches differ from ours in various ways, for instance, in~\cite{Bucher:1998mh} a background metric and an internal material metric are introduced into the model and its effects on the CMB were investigated. Dark matter models inspired by superfluidity were studied in~\cite{Berezhiani:2015bqa,Berezhiani:2015pia}.

The main idea of our approach, on the other hand, is to reinterpret the term $g^{\mu\nu}R_{\mu\nu}$ in the Einstein-Hilbert action as the isotropic limit of a more general theory. This approach is particularly natural in the context of the teleparallel equivalent of general relativity~\cite{Boehmer:2014jsa}. The inverse $g^{\mu\nu}$ is seen as the rank 2 isotropic tensor. This motivates an action based on the term $C^{\mu\nu}R_{\mu\nu}$ where $C^{\mu\nu}$ is a ``material'' tensor encoding the information about the internal degrees of freedom of the spacetime vacuum. The action of this theory is then given by
\begin{align}
  S = \int d^4x\sqrt{-g}\left(C^{\mu\nu}R_{\mu\nu} + \mathcal{L}_m\right) \,.
  \label{eq:action}
\end{align}
The metric remains the only dynamical degree of freedom in this theory and we only consider variations of the action with respect to $g_{\mu\nu}$ and the matter degrees of freedom. The material tensor is kinematical. 

This approach to modifying the Einstein-Hilbert action follows on from the ideas put forward by Brans and Dicke~\cite{Brans:1961sx}. They allowed for the gravitational constant to vary in space and time, thereby introducing an additional scalar degree of freedom which was treated dynamically. The Brans-Dicke model is partially contained in our approach when we choose $C^{\mu\nu} = \phi(x^{\alpha}) g^{\mu\nu}$. The main difference is that $\phi(x^{\alpha})$ is a dynamical degree of freedom while we keep $C^{\mu\nu}$ kinematical. Our approach is a natural generalisation allowing for tensorial modifications of the action. In other words, we allow for the gravitational field to be anisotropic in general. When compared to other modifications of General Relativity, our model is relatively harmless. The field equations of our model are still second order, no additional fields are introduced, local Lorentz invariance is not broken etc. One should think of the macroscopic form of Maxwell's equations where constitutive equations define the form of the dielectric tensor $\mathbf{D} = \varepsilon \mathbf{E}$, or using the index notation $D_i = \varepsilon_i{}^j E_j$. In general $\varepsilon_i{}^j$ is a rank 2 (kinematical) tensor which for an isotropic medium is given by $\varepsilon_i{}^j = \epsilon \delta^i_j$ where $\epsilon$ is the usual dielectric constant. We are extending this idea to general relativity. In the isotropic limit when we choose $C^{\mu\nu} = (c^4/16\pi G) g^{\mu\nu}$ we recover general relativity. It is interesting to note that similar actions to~(\ref{eq:action}) have been considered in Lorentz violating theories of gravity~\cite{Tasson:2014dfa}, however, they are motivated very differently. It is also interesting to note that~\cite{Verlinde:2016toy} also explored relations in the context of linear elasticity. This resulted in an expressions which relates the baryonic energy density to the dark matter energy density. 

The variation with respect to $g_{\mu\nu}$ gives the following gravitational field equations
\begin{multline}
  \Sigma^{\mu\nu\alpha\beta}R_{\mu\nu}-\frac{1}{2}g^{\alpha\beta}C^{\mu\nu}R_{\mu\nu}+\frac{1}{2} \Box C^{\alpha\beta}\\+\frac{1}{2}g^{\alpha\beta}\nabla_{\mu}\nabla_{\nu}C^{\mu\nu}-\nabla_{\mu}\nabla^{(\alpha}C^{\beta)\mu}=\frac{8\pi G}{c^4} T^{\alpha\beta} \,,
\label{eq:fullfieldeq}
\end{multline}
where $\Sigma$ is defined as
\begin{align}
  \Sigma^{\mu\nu\alpha\beta} = -\frac{\delta C^{\mu\nu}}{\delta g_{\alpha\beta}} \,.
\end{align}
One can verify that the choice $C^{\mu\nu} = g^{\mu\nu}$ will yield general relativity. Despite its slightly unusual form, this theory has some neat features. The field equations are of second order and the usual energy-momentum conservation equation holds due to Noether's theorem. This leads to an additional consistency equation 
\begin{align}
  J^{\beta} = 0 \,,
  \label{eqnj2}
\end{align}
where $J^{\beta}$ is given by the covariant derivative of the left-hand side of~(\ref{eq:fullfieldeq}) which reads
\begin{multline}
  J^{\beta} = \nabla_{\alpha} \Sigma^{\mu\nu\alpha\beta} R_{\mu\nu}
  +\Sigma^{\mu\nu\alpha\beta} \nabla_{\alpha} R_{\mu\nu}
  -\frac{1}{2} g^{\alpha\beta} R_{\mu\nu} \nabla_{\alpha} C^{\mu\nu} \\
  -C^{\alpha\sigma} \nabla_{\alpha} R^{\beta}{}_{\sigma}
  -R^{\beta}{}_{\sigma} \nabla_{\alpha} C^{\alpha\sigma} \,.
  \label{eqnj}
\end{multline}
One can also show that a Schwarzschild like solution exists~\cite{Boehmer:2013ss}.

In the following we will investigate the Newtonian limit of this theory by following the standard techniques~\cite{Will:2014kxa} of expanding the field equations around Minkowski space.

\section{Expansion around Minkowski space}

We want to expand the spacetime metric $g_{\mu\nu}$ around the Minkowski metric $\eta_{\mu\nu}=\mbox{diag}(-1,+1,+1,+1)$ and consider $C^{\mu\nu}$ to be nearly isotropic, i.e.~to differ from $g^{\mu\nu}$ only by a small amount. Therefore we will write
\begin{align}
  g_{\mu\nu} &= \eta_{\mu\nu}+h_{\mu\nu}\,,
  \nonumber \\
  C^{\mu \nu} &= g^{\mu\nu}-\varepsilon^{\mu\nu} = \eta^{\mu\nu}-h^{\mu\nu}-\varepsilon^{\mu\nu} \,.
  \label{eq:expansions}
\end{align}
The first relation in~(\ref{eq:expansions}) is nothing but the usual weak field limit of general relativity and we assume $|h_{\mu\nu}| \ll 1$ in order for this to be well-posed. The second linearisation in (\ref{eq:expansions}) corresponds to the assumption that $C^{\mu\nu}$ differs from isotropy, i.e.~from $g^{\mu\nu}$ corresponding to General Relativity (GR), just by a small amount $\varepsilon^{\mu\nu}$. In order for this expansion to be consistent we must assume $|\varepsilon^{\mu\nu}|\ll 1$. 

The minus signs in the decomposition of $C^{\mu\nu}$ in~(\ref{eq:expansions}) have been taken such that $C_{\mu\nu}=g_{\mu\nu}+\varepsilon_{\mu\nu}$ at first order in $\varepsilon^{\mu\nu}$. In other words, if we were to choose $C^{\mu \nu} = g^{\mu\nu}$, then we would have that $C^{\mu \nu} = \eta^{\mu\nu}-h^{\mu\nu}$ in first order, which is precisely our choice of signs.

This theory now depends on two small quantities, namely $h^{\mu\nu}$ and $\varepsilon^{\mu\nu}$ and there are no {\it a priori} reasons why one should be smaller than the other. However, experiments and observations at the Solar System scales show no trace of this anisotropy. The gravitational field around the Sun is spherical as opposed to ellipsoidal, say. Thus, we expect the effects of anisotropy perturbation $\varepsilon^{\mu\nu}$ to be smaller than the ones due to the metric perturbation $h_{\mu\nu}$. However, as we will see, even if the two magnitudes are comparable the phenomenology at small distances will not be changed. For this reason we will consider $|\varepsilon^{\mu\nu}|$ to be of the same order of $|h^{\mu\nu}|$ neglecting terms of $O(h^2)$, $O(\varepsilon^2)$ and $O(h\varepsilon)$. The next quantity in the field equations we need to consider is $\Sigma^{\mu\nu\alpha\beta}$.

We will assume that
\begin{align}
  \frac{\delta\varepsilon^{\mu\nu}}{\delta g_{\alpha\beta}} \sim O(h) \,, 
  \label{sigma}
\end{align}
which in turns implies $\Sigma^{\mu\nu\alpha\beta}=\eta^{\mu\alpha}\eta^{\nu\beta}+O(h)$. This is consistent with the Newtonian limit in GR. We are now ready to expand the field equations.

Expanding~(\ref{eq:fullfieldeq}) term by term to the required order gives the linearised field equations
\begin{multline}
  -\frac{1}{2}(\Box\overline{h}^{\alpha\beta}+\eta^{\alpha\beta}\partial_\mu\partial_\nu\overline{h}^{\mu\nu}-2\partial_\mu\partial^{(\alpha}\overline{h}^{\beta)\mu}) \\
  -\frac{1}{2}(\Box\varepsilon^{\alpha\beta}+\eta^{\alpha\beta}\partial_{\mu}\partial_{\nu}\varepsilon^{\mu\nu}-2\partial_{\mu}\partial^{(\alpha}\varepsilon^{\beta)\mu}) = 
  \frac{8\pi G}{c^4} T^{\alpha\beta} \,,
  \label{field}
\end{multline}
where we have defined the following tensor operations
\begin{align}
 \overline{S}_{\mu \nu} &= S_{\mu \nu}-\frac{1}{2}\eta_{\mu \nu}S \,,
 \nonumber \\
 S &= \eta_{\mu \nu}S^{\mu \nu} \,. 
\end{align} 

As is usual when linearising gravitational field equations, we work in the harmonic gauge $\partial_\mu \overline{h}^{\mu\nu}=0$ meaning that~(\ref{field}) simplifies to
\begin{align}
  - \frac{1}{2}\Box\overline{h}^{\alpha\beta} - \frac{1}{2}\Box\varepsilon^{\alpha\beta}
  - \frac{1}{2}\eta^{\alpha\beta}\partial_{\mu}\partial_{\nu}\varepsilon^{\mu\nu}
  + \partial_{\mu}\partial^{(\alpha}\varepsilon^{\beta)\mu}
  = \frac{8\pi G}{c^4} T^{\alpha\beta} \,.
  \label{field2}
\end{align}

We also note that the consistency equation~(\ref{eqnj2}) 
\begin{multline}
  J^{\beta} = \nabla_{\alpha} \Sigma^{\mu\nu\alpha\beta} R_{\mu\nu}
  +\Sigma^{\mu\nu\alpha\beta} \nabla_{\alpha} R_{\mu\nu}
  -\frac{1}{2} g^{\alpha\beta} R_{\mu\nu} \nabla_{\alpha} C^{\mu\nu} \\
  -C^{\alpha\sigma} \nabla_{\alpha} R^{\beta}{}_{\sigma}
  -R^{\beta}{}_{\sigma} \nabla_{\alpha} C^{\alpha\sigma} = 0 \,,
\end{multline} 
is automatically satisfied in this linear approximation. This is indeed expected as the theory should reduce to GR without additional constraints. 

\section{The Newtonian limit}

Next we wish to consider the linearised field equations in the slow moving case. We assume that $\partial_{0}h_{\mu\nu}$ and $\partial_{0}\varepsilon_{\mu\nu}$ can be neglected. Then the $(0,0)$-component of~(\ref{field2}) becomes
\begin{align}
  \nabla^2 \overline{h}^{00}+\nabla^2\varepsilon^{00}
  - \partial_{m}\partial_{n}\varepsilon^{mn} = 
  -\frac{16 \pi G}{c^4}T^{00} \,,
  \label{00}
\end{align}
and the $(i,j)$-components become
\begin{align}
  \nabla^2\overline{h}^{ij} + \nabla^2\varepsilon^{ij}
  - 2\partial_m\partial^{(i}\varepsilon^{j)m} + \delta^{ij}\partial_m\partial_n\varepsilon^{mn} = 
  - \frac{16\pi G}{c^4}T^{ij} \,.
  \label{ij}
\end{align}
By taking the 3 dimensional trace of the $(i,j)$-field equations~(\ref{ij}), we get
\begin{align}
  \nabla^2 \overline{h}^{ii}+\nabla^2\varepsilon^{ii} + 
  \partial_m\partial_n\varepsilon^{mn} = 
  -\frac{16 \pi G}{c^4}T^{ii} \,.
  \label{trace}
\end{align}
We can now add equations~(\ref{00}) and~(\ref{trace}) and arrive at
\begin{align}
  \nabla^2(\overline{h}^{00}+\overline{h}^{ii}+\varepsilon^{00}+\varepsilon^{ii}) = 
  -\frac{16 \pi G}{c^4}(T^{00}+T^{ii}) \,.
  \label{h}
\end{align}
Following the weak field limit approach used in General Relativity, we set $T^{00}=c^2\rho$ and $T^{ii}=0$ and find
\begin{align}
  \nabla^2(\overline{h}^{00}+\overline{h}^{ii}+\varepsilon^{00}+\varepsilon^{ii})
  = -\frac{16 \pi G}{c^2}\rho \,.
\label{h:new1}
\end{align}
This is our first significant result in the Newtonian limit of this theory. It should be noted that the quantities $h$ and $\varepsilon$ are both dimensionless.

Let us define
\begin{align}
  \varphi = -\frac{1}{4}(\overline{h}^{00}+\overline{h}^{ii}+\varepsilon^{00}+\varepsilon^{ii}) \,,
  \label{h2}
\end{align}
then $\varphi$ satisfies Poisson's equation $\nabla^2\varphi=\frac{4\pi G}{c^2}\rho$. One can verify that all quantities involved have the correct physical units. Note that both $h^{\mu\nu}$ and $\varepsilon^{\mu\nu}$ appear in the definition of the potential~(\ref{h2}). 

In the Newtonian limit of General Relativity, the field equations allow us to deduce $|\overline{h}^{00}| \gg |\overline{h}^{0i}|\gg|\overline{h}^{ij}|$. Assuming this here is equivalent to assuming $|\varepsilon^{00}| \gg |\varepsilon^{0i}| \gg |\varepsilon^{ij}|$, which we will assume from now on. In other words, if we assume $|\varepsilon^{00}| \gg |\varepsilon^{0i}| \gg |\varepsilon^{ij}|$, which naively corresponds to assuming that the ``sources'' due to the spacetime vacuum are slow moving, then from the field equations we obtain $|\overline{h}^{00}| \gg |\overline{h}^{0i}| \gg |\overline{h}^{ij}|$. This gives $\overline{h}^{00}=-\overline{h}=h$ and we find
\begin{align}
  h^{00} &= \frac{1}{2}h = -2\varphi-\frac{1}{2}\varepsilon^{00} \,, \\
  h^{11} &= h^{22} = h^{33} = -2\varphi-\frac{1}{2}\varepsilon^{00} \,.
\end{align}

In the following we analyse the far field of a stationary source. We set $T^{\mu \nu}=0$ far away from the source, so that~(\ref{h}) becomes
\begin{align}
  \nabla^2(\overline{h}^{00}+\varepsilon^{00}) = 0 \,,
\end{align}
which has the standard solution
\begin{align}
  \overline{h}^{00} + \varepsilon^{00} = \frac{K}{r}+O(r^{-2}) \,.
\end{align} 
Here $K$ is a constant of integration with units length. Comparison with equation~(\ref{h2}) yields
\begin{align}
  \varphi = -\frac{1}{4}(\overline{h}^{00}+\varepsilon^{00})
  = -\frac{GM}{c^2 r}+O(r^{-2}) \,,
  \label{eq:001}
\end{align}
which suggests that we should identify $K=4GM/c^2$.  

Thus far from a stationary source the spacetime metric is given by
\begin{align}
  ds^2 = &-[1-\frac{2GM}{c^2 }\frac{1}{r}+\frac{1}{2}\varepsilon^{00}+O(r^{-2})]c^2dt^2
  \nonumber\\
  &+[1+\frac{2GM}{c^2 }\frac{1}{r}-\frac{1}{2}\varepsilon^{00}+O(r^{-2})](dx^2+dy^2+dz^2).
\end{align}
We note that in both metric components the sign of $\varepsilon^{00}$ is opposite to the sign of the mass term. There is also a factor 4 difference and $\varepsilon^{00}$ is dimensionless. In order to change all this it turns out to be convenient to define
\begin{align}
  \sigma = -\frac{c^2\varepsilon^{00}}{4}
\end{align}
so that $\sigma$ has units of velocity squared. The metric component now is `symmetric' in the sense that both terms now have the same sign. Let us finally define
\begin{align}
  \frac{\bar{G}M}{r} = \frac{GM}{r} + \sigma
  \label{effG}
\end{align}
which we can view as the effective gravitational potential. Therefore, we can write the metric as
\begin{align}
  ds^2 = &-[1-\frac{2\bar{G}M}{c^2 }\frac{1}{r}+O(r^{-2})]c^2 dt^2 
  \nonumber \\ 
  &+[1+\frac{2\bar{G}M}{c^2}\frac{1}{r}+O(r^{-2})](dx^2+dy^2+dz^2) 
  \label{metric}
\end{align}
This can be neatly interpreted as a varying gravitational constant. Provided that $\bar{G}$ is approximately constant in the solar system, this solution will pass the three classical tests of general relativity; see Sec.~\ref{sec:NFW_profile} below for an explicit example.

Our interpretation of~(\ref{effG}) as the effective gravitational potential can be confirmed by considering the Newtonian limit of metric~(\ref{metric}). For this we assume  $\sigma$ to be a function of $r$ only and take the relevant limit $c \rightarrow \infty$. The only non-vanishing Christoffel symbol (besides the terms due to spherical symmetry) is given by 
\begin{align}
  \Gamma_{tt}^{r} [c \rightarrow \infty] = \frac{GM}{r^2}-\sigma'
  \label{eq:christoffel}
\end{align}
which is interpreted as the effective gravitational force, and matches the interpretation~(\ref{effG}). This is not too surprising as our theory can be seen as a generalisation of Brans-Dicke theory, see~\cite{Boehmer:2013ss}.

\section{Dark matter density profiles}

\subsection{The Navarro-Frenk-White profile}
\label{sec:NFW_profile}

As a first approximation we assume all the baryonic matter to be concentrated at the centre of the galaxy implying that we can effectively consider spherical symmetry. Of course corrections must be taken into account for applications to realistic galaxies: the baryonic matter in the outer parts of the galaxy will affect the results and the real form of a galaxy certainly do not respect spherical symmetry. However the scope of the present work is only to show that weak field limit applications of the theory advanced in~\cite{Boehmer:2013ss} can provide interesting features capable of mimicking dark matter at galactic scales. A complete treatment for realistic galaxies and a comparison with observational data is outside the objectives of the present analysis.

In the following we examine the gravitational rotation curves induced by this metric assuming that $\sigma$ is a function of the radius only. It is well known, see e.g.~\cite{Matos:2000ki}, that the tangential velocity of a test particle in a spherically symmetric metric is given by
\begin{align}
  \frac{v_{\rm tg}^2}{r^2} = \frac{c^2}{2r} \frac{d}{dr} \log g_{tt} \,.
  \label{grv}
\end{align}
Hence, using our metric~(\ref{metric}), we find for the tangential velocity
\begin{align}
  \frac{v_{\rm tg}^2}{r^2} &= \frac{c^2}{2r}
  \frac{(1-\frac{2\bar{G}M}{c^2 }\frac{1}{r})'}{1-\frac{2\bar{G}M}{c^2 }\frac{1}{r}}
  =\frac{1}{2r}\frac{(\frac{2GM}{r^2}-2\sigma')}{1-\frac{2GM}{c^2r}-2\frac{\sigma}{c^2}} 
  \nonumber \\
  &= \frac{1}{r} \Bigl(\frac{GM}{r^2}-\sigma'\Bigr)
  \Bigl(1-\frac{2GM}{c^2r}-2\frac{\sigma}{c^2}\Bigr)^{-1} \,.
\end{align}
By Taylor expanding the denominator, ignoring terms of $O(h^2)$, $O(h\varepsilon)$ and $O(\varepsilon^2)$, we find the following expression for the tangential velocity 
\begin{align}
  \frac{v_{\rm tg}^2}{r^2}=\frac{GM}{r^3}-\frac{\sigma'}{r} \,.
\end{align}
These assumptions are equivalent to saying $2GM/c^2 \ll r$ and $\sigma \ll c^2$. Equivalently, we could have Taylor expanded in $1/c^2$; compare with (\ref{eq:christoffel}).

Let us interpret the mass parameter $M$ as the baryonic mass, then we can write the tangential velocity as 
\begin{align}
  \frac{v_{\rm tg}^2}{r^2}=\frac{4\pi G}{3}\rho_{\rm baryonic}-\frac{\sigma'}{r} \,,
  \label{eq:vtg_with_sigmaprime}
\end{align}
and introduce an effective density $\rho_{\rm eff}(r)$ given by
\begin{align}
  \rho_{\rm eff} = \rho_{\rm baryonic} - \frac{3\sigma'}{4\pi Gr} \,.
  \label{eq:002}
\end{align}
It is important to note that $\rho_{\rm eff}$ is in fact not singular at the origin: We are working in the weak field limit which means our approximation is not valid for small values of $r$ where the gravitational field is strong. As such, we are not allowed to consider this approximation for small radii. 

In order to make a theoretical prediction, we need to specify the form of $\sigma$. The one thing we really know about gravity is that the Newtonian potential is inversely proportional to the radius, and that it works pretty well! Therefore, one of the simplest ways to parametrise $\sigma$ is to assume that $\sigma$ itself should be inversely proportional to the radius and to write
\begin{align}
  \sigma = \frac{G \mathcal{M}}{R_{\sigma} + r} = 
  \beta \frac{GM}{R_{\sigma} + r} \,,
  \label{signfw}
\end{align}
where $\mathcal{M}=M\beta$ is a constant with units mass and $\beta$ is dimensionless, its relevance becomes clear shortly.
It turns out that such a choice for $\sigma$ gives rise to the Navarro-Frenk-White (NFW) profile of dark matter halos~\cite{Navarro:1995iw}. We have
\begin{align}
  \rho =\frac{\rho_0}{\frac{r}{R_s}(1+\frac{r}{R_s})^2}\,,
  \label{rhonfw}
\end{align}
if we identify $R_{\sigma} = R_{s}$ and $\mathcal{M} = \rho_0 V = 4\pi \rho_0 R_s^3/3$. This is a quite remarkable result. By allowing the additional structure to vary like the gravitational potential we arrive at a somewhat natural explanation to flat galactic rotation curves (or dark matter) and are also able to give a good justification of the Navarro-Frenk-White profile.

The radius $R_{\sigma}$ is a constant which essentially determines at what distances the Newtonian laws are modified. We require $\sigma$ to be relevant only at galactic scales and to give no contributions at Solar System distances. At distances $r \ll R_\sigma$ we have $\sigma \simeq \sigma_{0} = \mathrm{constant}$ which implies no departures from the Newtonian dynamics on Solar System scales, see~(\ref{eq:002}). Recalling the metric (\ref{metric}) we can compute the post-Newtonian parameter $\gamma$ and write
\begin{align}
  2\gamma \frac{G M}{c^2 r} = 2 \frac{\bar{G} M}{c^2 r} = \frac{2}{c^2}\left(\frac{GM}{r}+\sigma\right) = 
  2\frac{GM}{c^2r}\left(1+\frac{\sigma r}{G M} \right)\,,
\end{align}
from which we can deduce that $\gamma-1$ is given by
\begin{align}
  \gamma -1 = \frac{\sigma r}{G M}\,.
\end{align}
Next, using the parametrisation (\ref{signfw}) we arrive at
\begin{align}
  \gamma -1 = \frac{\beta r}{R_{\sigma}}\left(1+\frac{r}{R_{\sigma}}\right)^{-1} \approx 
  \frac{\beta r}{R_{\sigma}}\,,
\end{align}
where we assumed that $r/R_\sigma \ll 1$ at solar system scales.

The Cassini bound constrains deviations on the post-Newtonian parameter $\gamma -1$ from zero, the current level \cite{Will:2014kxa} is $\gamma-1 < 10^{-5}$. We can make a rough order of magnitude estimate on the possible values of $R_\sigma$ allowed by this observation. This results in the simple bound $\beta r/R_\sigma < 10^{-5}$. Taking the radius $r$ to be roughly solar system distances, approximately $10^{15} {\rm m}$, one finds
\begin{align}
  \beta < 10^{-20} {\rm m}^{-1} \times R_\sigma \,.
\end{align}
For the NFW profile, fits from numerical simulations to the observed rotation curves imply $R_\sigma = R_s \gtrsim 10^{21} \,{\rm m}$ $(\sim 100\, {\rm kpc})$, see \cite{Navarro:1995iw}. Consequently, one can thus safely identify $R_\sigma$ with $R_s$ as long as $\beta$ satisfies the bound $\beta < 10$.

To have an idea of the possible values of $\beta$, we must recall that in deriving these results, we made the assumptions that $\varepsilon$ and $h$ are of the same order. This is equivalent to the parameter $\beta$ being of the order one, $\beta \simeq O(1)$. We must check this assumption is indeed valid. This is a crucial check to the validity of our result. So far, we have derived and solved the modified field equations making a variety of assumptions to treat certain quantities as being small. Next, we will see that our results are consistent with various galaxies. Now, we are using best fit estimates of the NFW profile parameters $R_{s}$ and $\rho_0$ from the 19 galaxies spanning four orders of magnitude of mass given in~\cite{Navarro:1995iw}. We find estimates of $\beta$ in the range $1 < \beta < 5$. These galaxies provide strong evidence that $\beta$ is indeed of $O(1)$. This in turn justifies our assumptions and shows the NFW profile can appear somewhat naturally in our theory. Note that these values of $\beta$ only satisfy the above Solar System constraints by one order of magnitude. This implies that in principle the theory could be tested and possibly falsified by future experiments.

It is well known that the NFW profile has some shortcomings and does not accurately describe the rotation curves of many galaxies~\cite{deBlok:2009sp,Castignani:2012sr}. We will therefore consider other density profiles in the context of our model. 

\subsection{Burkert density profile}

It is important to emphasise that our model contains an additional kinematical degree of freedom in the form of the material tensor $C^{\mu\nu}$. As there are no constraints as to how this is chosen, in principle, our model is able to reproduce any given velocity profile. The same holds true for general relativity where we could use~(\ref{grv}) to find the metric function $g_{tt}$ and then use the Einstein field equations to determine the remaining components of the metric. This approach tends to result in metrics with singularities. In our model, the dark matter density profile is the second term in~(\ref{eq:002}) which gives
\begin{align}
  \rho_{\rm dm} = - \frac{3 \sigma'}{4 \pi G r}.
  \label{rhodm1}
\end{align}
This allows us to find $\sigma$ by integration from a given dark matter profile 
\begin{align}
  \sigma = \sigma_0 - \int \frac{4\pi G r}{3} \rho_{\rm dm}(r)\, dr,
  \label{n1}
\end{align}
where $\sigma_0$ is the constant of integration. This determines $\sigma$ for given dark matter profile $\rho_{\rm dm}(r)$ which in turn is related to the velocity profile. Thus we are able to find the forms of $\sigma$ for various galaxies, including the outer parts of the galaxy where for instance the NFW profile no longer matches observations well~\cite{deBlok:2009sp,Castignani:2012sr}. However, there is little predictive power in this approach unless one can find a `universal' function $\sigma$, depending on a few constants, which matches galactic rotation data well for a large number of galaxies. In this sense it is a matter of taste whether one prefers to determine $\rho_{\rm dm}$ or $\sigma$ from observations.

For concreteness let us consider the Burkert profile~\cite{Burkert:1995yz} whose density profile is
\begin{align}
  \rho_{\rm dm} = \frac{\rho_0 r_0^3}{(r+r_0)(r^2+r_0^2)}
  \label{rhobur}
\end{align}
where $\rho_0$ is the central density and $r_0$ is a scaling constant. The Burkert profile accurately describes the observed rotation curves of dwarf galaxies, which are known to be dark matter dominated, and in contrast to the NFW profile the Burkert profile has a central core.

Using~(\ref{n1}), our model can replicate such a dark matter profile provided $\sigma$ is of the form
\begin{align}
  \sigma = \sigma_0 - \frac{1}{4} \frac{G\mathcal{M}}{r_0}
  \left(\arctan(r/r_0) - \log\frac{(1+r/r_0)^2}{1+r^2/r_0^2}\right)
  \label{sigbur}
\end{align}
where we defined the quantity $\mathcal{M} = 4\pi r_0^3 \rho_0 /3$, as above. We note that $\sigma$ approaches a constant value when  $r \gg r_0$ which corresponds to the velocity approaching zero.

An interesting point can be made here. When comparing the NFW profile~(\ref{rhonfw}) with the Burkert profile~(\ref{rhobur}), they appear to be `similar' in the sense that their functional forms do not differ significantly. However, this cannot be said for their corresponding forms of $\sigma$, compare~(\ref{signfw}) with~(\ref{sigbur}), they are very different. The reason for this comes mainly from the integration in~(\ref{n1}). It could therefore be of interest to study~(\ref{n1}) in some detail for given density profiles of a variety of galaxies.

\section{Discussion}

A simple form for the function $\sigma(r)$ would be a polynomial function in the radius $r$. Interestingly, such a simple choice is in good agreement with previous studies. For instance in~\cite{Martins:2007uf} a power law correction to the Newtonian potential was considered which would also correspond to a power law form of $\sigma$. A similar result was found by~\cite{Mannheim:2012qw} where the corrections to the baryonic velocity profile were linear and quadratic in the distance from the centre. This would again correspond to a polynomial form for the material  function $\sigma$. The velocity profile derived in~\cite{Boehmer:2007um} led to a trigonometric function in the radius, however, this can also be well approximated to be a polynomial for small distances from the centre of the galaxy. We would also like to mention to logarithmic correction suggested in~\cite{Fabris:2007df} which would also correspond to a logarithmic form of $\sigma$.

Within our effective framework we can reinterpret different approaches to dark matter by providing an effective description which includes the phenomenology of many of those models. For instance, in Modified Newtonian Dynamics, galactic rotation curves are asymptotically flat~\cite{Milgrom:1983pn,Milgrom:1983ca,Sanders:2002pf}, see also~\cite{Bekenstein:2004ne,CervantesCota:2009my} for a relativistic formulation and applications. Following Eq.~(\ref{eq:vtg_with_sigmaprime}) this constraints the asymptotic form of $\sigma'$, namely MOND-like behaviour requires a decay of the form $\sigma' \sim 1/r$. This means we require logarithmic terms in $\sigma$ like those present in~(\ref{sigbur}). It is perhaps unsurprising that Brans-Dicke theories with suitably chosen potentials can also explain rotation curves, see for instance~\cite{Gessner:1992flm,Burrage:2016yjm}.

Another example is the Emergent Gravity Paradigm~\cite{Verlinde:2016toy} where the dark matter energy density is related to the baryonic one. In our description this could be achieved by assuming $\sigma$ to be a function of $\rho_{\rm baryonic}$. Note that this would require the additional assumption in our model as the `material' spacetime structure due to $\sigma$ would be related to the matter. Finally, we would like to mention the modified gravity model proposed by Moffat~\cite{Moffat:2013sja} which also yields flattened galactic rotation curves. 
 
A similar observation can be made for the Mass-Discrepancy-Acceleration Relation~\cite{1990A&ARv,2016PhRvL.117t1101M,Lelli:2017vgz} which shows a simple relationship between the radial acceleration due to dark matter and baryonic matter. In our model, the gravitational force or acceleration, see Eq.~(\ref{eq:christoffel}), depends directly on $\sigma'$ suggesting a relationship between $\sigma'$ and the acceleration due to the baryonic matter. An interesting issue to investigate in the future would be to find a phenomenological parametrisation of $\sigma$, depending on various constants and on the baryonic matter profile, which allows one to describe the discussed models explicitly. 

\section{Conclusion}

We studied the Newtonian limit of a modification of General Relativity which is based on ideas from Continuum Mechanics. By expanding the metric about Minkowski space and by assuming small deviations from the isotropy of the gravitational force, we were able to formulate the Newtonian limit of the theory, equation~(\ref{h:new1}). We solved this equation finding the gravitational field far from a static and spherically symmetric source. The resulting metric can be interpreted as a modification of General Relativity with an effective gravitational constant. These results were then applied in the context of dark matter halos in galaxies. Assuming that the additional structure of spacetime $\sigma$ behaves like the gravitational potential led to the NFW profile. This assumption on $\sigma$ is equivalent to considering the Navarro-Frenk-White profile. We also considered our model in a more general context and discussed how the Burkert profile can be replicated in this setting, additionally we also discussed some dark matter profiles used in previous work and their relation to our model. 

\subsection*{Acknowledgements}
The authors would like the thank the referee for constructive feedback on the manuscript, and thank Tiberiu Harko for useful discussions. This project was supported in part by Scheme 4 grant 41220 of the London Mathematical Society. N.T.~acknowledge support from the Labex P2IO and an Enanched Eurotalents Fellowship.

\end{document}